\documentclass[aps,prl,reprint,showpacs]{revtex4-1}

\usepackage{bm}
\usepackage{amssymb}
\usepackage[intlimits]{amsmath}
\usepackage{graphicx}
\usepackage{xcolor}
\definecolor{olive}{rgb}{0,0.5,0}
\definecolor{maroon}{rgb}{0.5,0,0}

\newcommand{\revision}{}

\begin{document}

\title{Rashba-coupling modelling for two-dimensional and high-order Rashba Hamiltonian for one-dimensional confined heavy holes} %Title of paper

\author{R. Cuan}
\email[E-mail me at: ]{rcuan@fisica.uh.cu}
\affiliation{Facultad de F\'{\i}sica. Universidad de La Habana, C.P.10400, La Habana, Cuba.}

\author{L. Diago-Cisneros}
\email[E-mail me at: ]{ldiago@fisica.uh.cu}
\affiliation{Facultad de F\'{\i}sica. Universidad de La Habana, C.P.10400, La Habana, Cuba.}
\affiliation{Departamento de F\'{\i}sica y Matem\'{a}ticas, Universidad Iberoamericana,  M\'exico D. F.}

\date{\today}

\begin{abstract}
Based on standard \textbf{k}$\cdot$\textbf{p} ($8 \times 8$) multiband Hamiltonian, we have deduced an explicit analytical expression for the Rashba-coupling parameter which clarifies its anomalous behavior for heavy holes ($hh$), gated in quasi-two-dimensional (Q2D) systems, by letting grow the density. Our modelling remarkable better agrees with experimental results in comparison with earlier theoretical models, while recovers the expected cubic dependence on the quasi-momentum. For quasi-one-dimensional (Q1D) $hh$ systems, we have formally derived an effective Rashba Hamiltonian with two competitive terms on the quasi-momentum, a linear term and a cubic one as predicted from suitable approximations to the Q2D scope. The Rashba-coupling parameters also behave anomalously and qualitatively support recent experiments in core/shell nanowires. Furthermore, they exhibit an essential asymptotic discontinuity in the low density regime as a function of the lateral confinement length. For $hh$, we present closed schemes to accurately quote the Rashba-coupling parameters both for the Q2D and Q1D systems, which become unprecedented for holes.
\end{abstract}


\maketitle %\maketitle must follow title, authors, abstract and \pacs

One of the biggest challenges in Spintronics \cite{Awschalom07} is to manipulate efficiently spin currents in semiconductors systems without external magnetic fields. Rashba spin-orbit interaction (SOI-R) \cite{Rashba84a,Rashba84b} is among the most promising mechanisms for achieve this \cite{Datta90,Pala04,Shelykh05,Leo13}. SOI-R  for the heavy holes ($hh$) case is different from that for electrons or light holes ($lh$). In quasi-two-dimensional (Q2D) systems, observed in experiments since 1984 \cite{Wieck84}, the SOI-R coupling Hamiltonian exhibits an atypical cubic dependence in the quasi-momentum ($\hat{k}^3$) obtained by Gerchikov and Subashiev \cite{Gerchikov92}, and Winkler \textit{et al.} \cite{Winkler02,BookWinkler03,Habib09} using group theory arguments.

The long data assumption that the induced SOI-R  spin splitting in Q2D systems at zero magnetic field rises with electric field, responsible for the inversion asymmetry of the confining potential, firmly validated for electrons \cite{Andrada94}, was definitely abated when the opposite was demonstrated for holes \cite{Winkler02}. Importantly, it has been reported, both numerically and experimentally, the SOI-R coupling parameter anomalous decreasing with a non-zero electric field for Q2D $hh$ in accumulation-layer-like single heterostructures \cite{Winkler02,BookWinkler03}. Recently, Habib \textit{et al.} \cite{Habib09} have shown interesting measurements, with a similar behavior at fixed density, but tuning the external electric field perpendicular to the Q2D gas. Notwithstanding these achievements, several features in the anomalous SOI-R coupling effect remains cumbersome and inspired us to address a complementary study.

On the other hand, despite the efforts focused to study SOI-R  in quasi-one-dimensional (Q1D) holes systems, due to their appealing applications and phenomenology  \cite{Quay10,Governale03,Zhang07,Cuan11,Chesi11,Hao10}, some relevant topics remains incomplete still or deserve further attention. SOI-R  model for Q1D-$hh$ systems suggested by Governale and Z\"{u}licke \cite{Governale03}, and intuitively based on  the behavior  of the Dresselhaus term for Q2D electrons \cite{Rashba88}, leads to a Hamiltonian linear in the quasi-momentum. However, Zhang and Xia \cite{Zhang07} suggested that additional competitive high-order terms may be needed to explain their finding regarding the spin splitting energy of $hh$ in cylindrical quantum wires under external electric fields. Likewise, from our calculation of $hh$ dispersion laws in Q1D systems \cite{Cuan11}, patterned in Q2D hole gases by repulsive bias, we demonstrated the existence of a cubic-in-the-quasi-momentum term. Recently, Chesi \textit{et al.} \cite{Chesi11} proposed a clever model, replacing the quasi-momentum operator along a constricted direction of a Q2D-$hh$ system by its expectation value. They obtained two competitive terms, a linear and a cubic one, that explain the anomalous sign of the spin polarization filtered by a quantum point contact \revision{(QPC)}, as observed in magnetic focusing experiments, as well as the crossing or anticrossing of spin-split levels depending on subband index and the magnetic field direction \cite{Chesi11,Nichele14}. For electrons, some theoretical studies argue that the SOI-R coupling parameter decreases with increasing wire confinement \cite{Andrada03}, however later experimental measurements \cite{Guzenko06} for widths ranging from 1.18~$\mu$m down to 210~nm  suggest the opposite. For $hh$, one follow that the factor associated to the linear term behaves as theoretically predicted for Q1D electrons, while the one associated to the cubic term do not depend on the wire width \cite{Chesi11}. Moreover, recent magneto-transport measurements in Ge/Si core/shell nanowires \cite{Hao10} show that SOI-R  strength of holes behaves anomalously, as in the Q2D case. We get motivated about the undoubted existence of competitor high-order terms on the quasi-momentum within the SOI-R  model for Q1D-$hh$ systems, and the lack of a formal treatment for the Q2D-to-Q1D transition pursuing an appropriate Q1D-$hh$ SOI-R  model that fulfill the issues presented above.

In this Letter, we revisit the SOI-R  modelling for Q2D-$hh$ systems and we manage to improve the development by Winkler \textit{et al.} \cite{Winkler02} We show that our representation of the SOI-R coupling parameter, support its subtle anomalous behavior, when one let the density grow. We have deduced this quantity explicitly, compare it with available experimental data as well as with other numerical simulations, and stress the remarkably better agreement we have found. A similar procedure allow us to derive a model Hamiltonian for SOI-R  in Q1D-$hh$ systems.

It is widely accepted that SOI-R  for the valence band (VB) is a second order effect. \revision{We are interested in quantum systems made of direct semiconductors with zinc-blende structure (point group $T_d$), in particular those of III-V and II-VI binary compounds.} Consequently, we start from the \textbf{k}$\cdot$\textbf{p} ($8\times 8$) Pidgeon-Brown \revision{(PB)} model in the spherical approximation \cite{Pidgeon66,Efros98}, which involves VB and the conduction band (CB). The inclusion of \revision{CB} is also supported by results \cite{BookWinkler03} which suggest that SOI-R  is strongest in narrow-gap semiconductors. Then, by considering proper confining potential, the Q2D or Q1D are focused.

To accurately describe a Q2D system of confined holes at the heterostructure interface, the Poisson and Schr\"{o}dinger equations have to be self-consistently solved. In the framework of our scheme, we model the heterostructure potential profile as a triangular well, thus $V(z) = eFz$, where $F=eN_{s}/(2\varepsilon\varepsilon_{0})$ \cite{Winkler02}. The present approximation had been successfully applied in the study of spectral properties of electrons \cite{Andrada94} as well as for hole \cite{Cuan11}.

Focusing on extracting from the Schr\"{o}dinger eight-component system\revision{\footnote{\revision{Here $\hat{\bm {H}}_{\text{PB}}$ is the Pidgeon-Brown Hamiltonian while $(\mathbb{I/O})_{8}$ stands for the eight-dimension identity-matrix/null-vector, respectively.}}}
\begin{equation}
 \left[\hat{\bm {H}}_{\text{PB}} + (V(z) - \mathcal{E})\mathbb{I }_{8}\right]\bm{\varPsi}(\bm r) = \mathbb{O}_{8},
\end{equation}
an effective Hamiltonian for describing the $hh$ states, we use the partition L\"{o}wdin scheme \cite{Lowdin51,BookWinkler03} \revision{up to second order  of perturbation
\begin{align}
  H_{mm'}^{(0)} & =  H_{mm'}^{0},\\
  H_{mm'}^{(1)} & = H'_{mm'},\\
  H_{mm'}^{(2)} & =  \dfrac{1}{2}\sum_lH'_{ml}H'_{lm'}\left[ \dfrac{1}{E_m-E_l} + \dfrac{1}{E_{m'}-E_l}\right],\label{eq:Hparc}
\end{align}
with $H_{mm'}^{0}$ and $H'_{mm'}$ being the diagonal and off-diagonal component of $\hat{\bm {H}}_{\text{PB}}$, respectively.

We are interested in the elements $H_{hh\uparrow,hh\downarrow}$ and $H_{hh\downarrow,hh\uparrow}$ ---$H_{3,6}$ and $H_{6,3}$ in $\hat{\bm {H}}_{\textsc{PB}}$ for the basis chosen in Ref. \cite{Efros98}--- which couple the heavy holes with different total angular momentum projection. Taking into account that $H_{mm'}^{(0)}$ and $H_{mm'}^{(1)}$ are both null \cite{Efros98}, after some simple algebra in the term (\ref{eq:Hparc}) we finally obtain
\begin{multline}\label{eq:H36d22}
H_{hh\uparrow,hh\downarrow}=\\
\dfrac{\langle\varphi_n^{hh}|L|\varphi_n^{lh}\rangle \langle\varphi_n^{lh}|M| \varphi_n^{hh}\rangle - \langle\varphi_n^{hh}|M|\varphi_n^{lh}\rangle \langle\varphi_n^{lh}|L| \varphi_n^{hh}\rangle}{E_n^{hh}-E_n^{lh}} \\
+\dfrac{\langle\varphi_n^{hh}|M|\varphi_n^{\revision{ss}}\rangle \langle\varphi_n^{\revision{ss}}|L| \varphi_n^{hh}\rangle - \langle\varphi_n^{hh}|L|\varphi_n^{\revision{ss}}\rangle \langle\varphi_n^{\revision{ss}}|M| \varphi_n^{hh}\rangle}{E_n^{hh}-E_n^{\revision{ss}}},
\end{multline}
and
\begin{multline}\label{eq:H63d22}
H_{hh\downarrow,hh\uparrow}=\\
\dfrac{\langle\varphi_n^{hh}|L^{*}|\varphi_n^{lh}\rangle \langle\varphi_n^{lh}|M^{*}| \varphi_n^{hh}\rangle - \langle\varphi_n^{hh}|M^{*}|\varphi_n^{lh}\rangle \langle\varphi_n^{lh}|L^{*}| \varphi_n^{hh}\rangle}{E_n^{hh}-E_n^{hh}} \\
+\\
\dfrac{\langle\varphi_n^{hh}|M^{*}|\varphi_n^{\revision{ss}}\rangle \langle\varphi_n^{\revision{ss}}|L^{*}| \varphi_n^{hh}\rangle - \langle\varphi_n^{hh}|L^{*}|\varphi_n^{\revision{ss}}\rangle \langle\varphi_n^{\revision{ss}}|M^{*}| \varphi_n^{hh}\rangle}{E_n^{hh}-E_n^{\revision{ss}}}.
\end{multline}
In the equations (\ref{eq:H36d22}) and (\ref{eq:H63d22}), $L$ and $M$ are terms of $\hat{\bm {H}}_{\text{PB}}$ given by
\begin{equation}
 L=\dfrac{-\sqrt{3}\hbar^{2}\gamma_s}{m_0}k_-\dfrac{d}{dz}, \qquad M=\dfrac{\sqrt{3}\hbar^{2}\gamma_s}{2m_0}k^2_-,
\end{equation}
where $\hat k_{\pm} = \hat k_{x}\pm i\hat k_{y}$, $\hbar$ is the Planck constant, $m_0$ is the bare electron mass and $\gamma_{s} = (2\gamma_{2}^{L} + 3\gamma_{3}^{L})/5-E_p/(6E_g)$, with $\gamma_{1,2,3}^{L}$ as the semi-empirical L\"{u}ttinger parameters, $E_p$ is related to the Kane parameter $P_K=-i(\hbar/m_0)\langle S|p_x|X\rangle$ through $E_p=(2m_0/\hbar^{2})P_K^{2}$, and $E_g$ is the energy gap. Likewise,
\begin{equation}\label{eq:airy}
  \varphi_n^{hh,lh,\revision{ss}}(z)=\sqrt{\dfrac{1}{z^{0}_{hh,lh,\revision{ss}}\mathcal{A}'^{2}(-c_n)}}\mathcal{A}\left(\dfrac{z}{z^{0}_{hh,lh,\revision{ss}}}-c_n\right),
\end{equation}
are the normalized solutions of the VB transversal states, for the diagonal part of $\hat{\bm {H}}_{\text{PB}}$ with a triangular quantum well \cite{Cuan11}, with $z^{0}_{hh,lh,\revision{ss}}=(2m^{*}_{hh,lh,\revision{ss}}eF/\hbar^{2})^{\frac{1}{3}}$, $m^{*}_{hh/lh/\revision{ss}}$ the effective mass parallel to the growth direction $z$, for the heavy-, light- and spin-splitted split-off holes: $m_0/(\gamma_1-2\gamma_s)$, $m_0/(\gamma_1+2\gamma_s)$ and $m_0/(\gamma_1)$, respectively. We represent by $\mathcal{A}$ one of the Airy functions, while $\mathcal{B}$ ---the other one--- diverges at $\infty$, being $c_{n} = 2.338, 4.088, 5.521,\dots$ its zeros. Furthermore, the transversal energy levels are given by
\begin{equation}
 \label{eq:En}
   E_{n}^{hh,lh} = c_{n}\dfrac{(\hbar eF)^{\frac{2}{3}}}{\sqrt[3]{2m^{*}_{hh,lh}}},\,\,
   E_{n}^{\revision{ss}} = c_{n}\dfrac{(\hbar eF)^{\frac{2}{3}}}{\sqrt[3]{2m^{*}_{\revision{ss}}}}+\varDelta_{SO},
\end{equation}
with $\varDelta_{SO}$ as the split-off energy.

Finally, the terms (\ref{eq:H36d22}) and (\ref{eq:H63d22}) lead us to recover the expected form of the Rashba model for the Q2D case \cite{Gerchikov92,Winkler02,BookWinkler03,Habib09}
\begin{equation}
 \label{eq:HR2D}
 \hat{\bm {H}}_{R-2D}^{hh} = \beta(\bm \sigma_{+}\hat k_{-}^{3} + \bm \sigma_{-}\hat k_{+}^{3}),
\end{equation}
and to derive a new expression for the SOI-R parameter $\beta$,
\begin{equation}
 \label{eq:betaQ2D}
  \beta_{n} = \dfrac{-3i\hbar^{4}\gamma_{s}^{2}}{m_{0}^{2}}
  \left(\dfrac{I^{hl}_{n}Z^{hl}_{n}}{E_{n}^{hh}-E_{n}^{lh}}- \dfrac{I^{hs}_{n}Z^{hs}_{n}}{E_{n}^{hh}-E_{n}^{\revision{ss}}}\right),
\end{equation}
where $\bm \sigma_{\pm} = \frac{1}{2}(\bm\sigma_{x} \pm i\bm\sigma_{y})$ and $\bm \sigma_{x,y}$ the Pauli matrixes, while $I_{n}^{hl/hs}$ and $Z_{n}^{hl/hs}$ are Airy function integrals of the form
\begin{equation}\label{eq:InZn}
 I_n^{hl,hs}=\left\langle\varphi_n^{hh}|\varphi_n^{lh,\revision{ss}} \right\rangle,\,\,\, Z_n^{hl/hs}=\left\langle\varphi_n^{hh}|\hat k_z|\varphi_n^{lh,\revision{ss}}\right\rangle.
\end{equation}

It is worthwhile to underline that the non-perturbated term (diagonal part) includes the electric field and we were able to solve it exactly, which is a differentiating advantage of the present approach respect to earlier models \cite{Winkler02}, which considered the electric field perturbatively and only in the third-order terms can be approximately obtained the none-zero off-diagonal matrix elements. We had disregarded the third-order perturbation term due to its contribution of about $0.2$ percent.
}

The deduced expressions \revision{(\ref{eq:airy}), (\ref{eq:En}) and (\ref{eq:InZn})} represent a reliable starting-point platform to quote SOI-R coupling parameter (\ref{eq:betaQ2D}) for specific semiconducting Q2D-$hh$ systems, playing for $hh$ the same role as expression (13) reported before for electrons \cite{Andrada94}. The completeness of these self-complementary exact formulae is unprecedented for holes, as far as we know.

Following expression (\ref{eq:betaQ2D}), the existence of an inverse  $\beta$ dependence on energy-levels (\ref{eq:En}) gaps for the VB states is observed. This suggests that VB levels separation, which is in principle an electric-field $F$ strength-assisted raising process and/or a charge-carrier concentration $N_{s}$ growing function, governs the charge carriers confinement. Furthermore, our result (\ref{eq:betaQ2D}), is consistent with the anomalous Rashba spin-splitting \cite{Winkler02,BookWinkler03,Habib09}.

\begin{figure}
\centering
\includegraphics[width=0.85\linewidth]{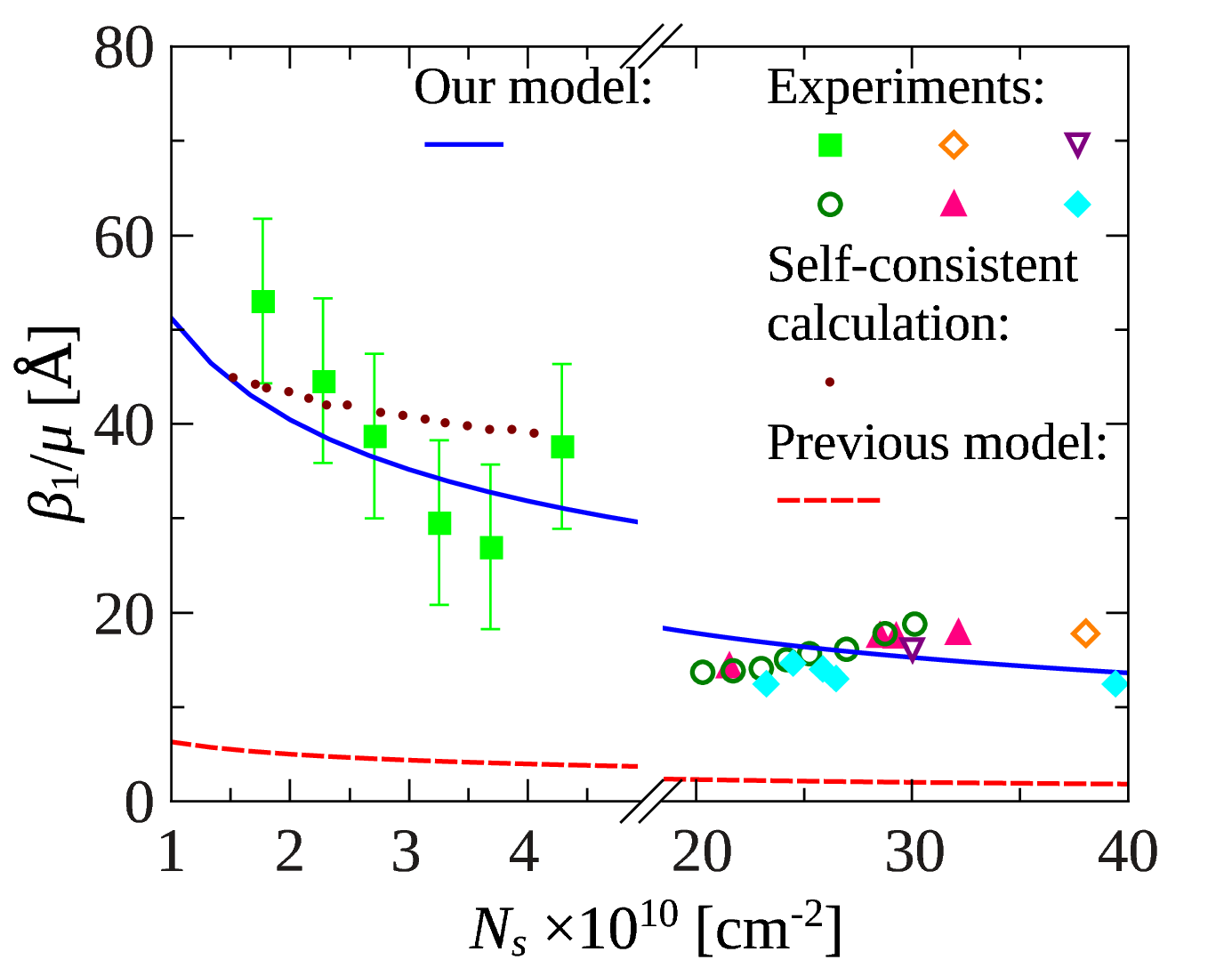}
\caption{(Color online) Q2D coupling parameter $\beta_{1}$ for the first $hh$ subband, taken after (\ref{eq:betaQ2D}), is shown (\textcolor{blue}{solid line}), and compared with previous analytical \cite{Habib09} (\textcolor{red}{dashed line}), experimental (symbols:  \textcolor{green}{$\blacksquare$} Ref.~\cite{Winkler02}, \textcolor{magenta}{$\blacktriangle$} Refs.~\cite{Grbic04,Grbic05}, {\Large \textcolor{olive}{$\circ$}} Ref.~\cite{GrbicPhD07}, \textcolor{orange}{$\Diamond$} Ref.~\cite{Grbic08}, \textcolor{violet}{$\triangledown$} Ref.~\cite{Grbic08a}, and \textcolor{cyan}{$\blacklozenge$} Ref.~\cite{Lu98}) and numerical \cite{Winkler02,BookWinkler03}  results (\textcolor{maroon}{dotted line}) based on a self-consistent method.}
\label{fig:GaAs-beta}
\end{figure}

Fig. \ref{fig:GaAs-beta} displays $\frac{\beta_{1}}{\mu_{h}}$, with $\mu_{h}=\hbar^{2}/2m_{hh-Q2D}^{*}$, as a function of  $N_{s}$ (solid line) following (\ref{eq:betaQ2D}). Besides, pursuing a validation of our modelling approach, a comparison with some relevant analytical \cite{Habib09}, experimental  \cite{Winkler02,BookWinkler03,Grbic05,GrbicPhD07,Grbic08,Grbic08a,Lu98} and numerical \cite{Winkler02,BookWinkler03} contributions, had been included. In the case of experimental measurements the samples are of the type: Al$_x$Ga$_{1-x}$As/GaAs, for several values of molar composition $x$ and acceptor doping. Before continuing, let us briefly mention some specific details, namely: (\textit{i}) The expression (\ref{eq:betaQ2D}) is ``autonomous" upon events outside the region where the Q2D gas is built and depend solely on GaAs layer's parameters. (\textit{ii}) In the case of Refs.~\cite{Grbic05,GrbicPhD07,Grbic08,Grbic08a,Lu98}, reports of carriers' concentration measurements with different spin polarizations ($N_{+},N_{-}$)  were taken, and had been considered an expression that relate them with $\beta_1$ \cite{Winkler02,BookWinkler03}
\begin{equation}
 \frac{\beta_{1}}{\mu_{h}} = \sqrt{\frac{2}{\pi}}\frac{N_{s} (\tilde{N}_{s}^{+} - \tilde{N}_{s}^{-}) + \varDelta N_s(\tilde{N}_{s}^{+} + \tilde{N}_{s}^{-})}{6N_{s}^{2} + 2\varDelta N_{s}^2},
\end{equation}
being $\tilde{N}_{s}^{\pm} = \sqrt{N_{s}\pm\varDelta N_{s}}$. Confronting our model (\ref{eq:betaQ2D}) to the above mentioned experimental data, general reasonable coincidence has been found. Notice the remarkably good agreement within the low density range [$1\times10^{10}-5\times 10^{10}$]$~cm^{-2}$, when our simulation (solid line) crosses almost all experimental-point uncertainty zones, following as well the overall experimental-data trend. For larger density section [$20\times10^{10}-40\times 10^{10}$]$~cm^{-2}$, we have achieved a better match for $\beta_{1}$ concerning its measured quantity, meanwhile a slight mismatch in tendency have been detected. Finally, we turn to previous numerical simulations \cite{Winkler02,BookWinkler03,Habib09}. The case of Refs.~ \cite{Winkler02,BookWinkler03}, represents a local very precise fitting (dotted line) as can be seen in the region [$2\times10^{10}-3\times 10^{10}$]$~cm^{-2}$. On the contrary, in the range [$3\times10^{10}-4\times 10^{10}$]$~cm^{-2}$, that numerical prediction noticeably departs from experiment. Globally speaking, we think that this calculation \cite{Winkler02,BookWinkler03} do not fully reproduce the overall tendency of the experimental points, in the range [$1\times10^{10}-5\times 10^{10}$]$~cm^{-2}$, which contrast with the finest adjustment exhibited by the solid curve (\ref{eq:betaQ2D}) in the same range. Worthwhile to underline that analytically quoted results (dashed line) of Ref.~\cite{Habib09}, are clearly far both from experimental points and from
the above discussed numerical modelling (including ours). However, the general trend is some how preserved, specially in the high density region.

\begin{figure}
\centering
\vspace{6mm}\includegraphics[width=0.85\linewidth]{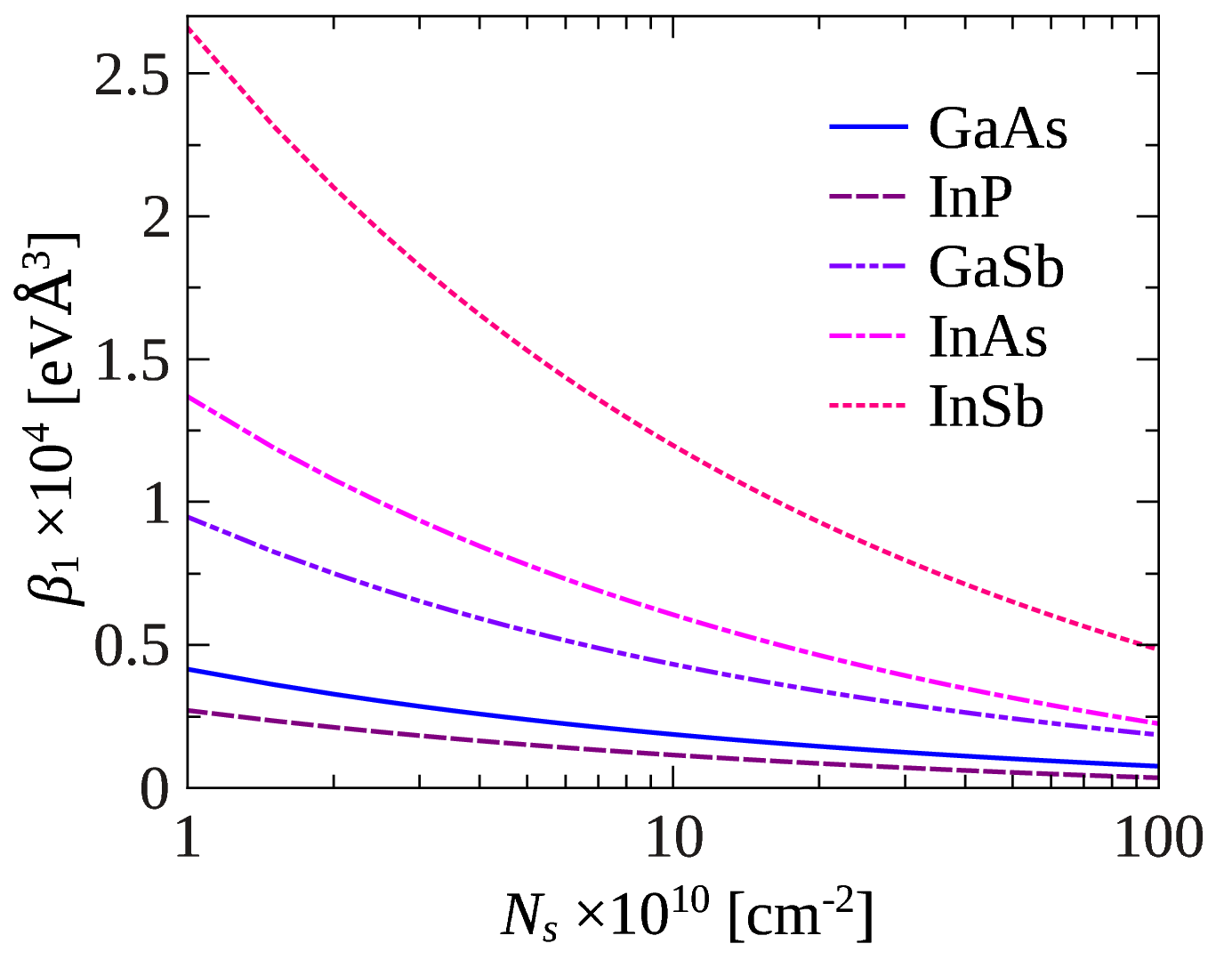}
\caption{(Color online) SOI-R coupling parameter $\beta_{1}$ as a function of $N_{s}$, for the lower $hh$ sub-band in different semiconductors.}
\label{fig:beta2d}
\end{figure}

Fig.\ref{fig:beta2d} plots  $\beta_{1}$ within typical density range [$10^{10}-10^{12}\,cm^{-1}$] and exercised several semiconductors with narrow: InSb ($0.235\,eV$, dashed-double/dotted line), InAs ($0.417\,eV$, dashed-dotted line); middle: GaSb ($0.812\,eV$, dotted line); and wide: InP ($1.4236\,eV$, dashed line), GaAs ($1.519\,eV$, solid line) energy gaps, respectively. One can easily observe all analyzed materials satisfy the predicted anomalous trend. Nonetheless, the expected \cite{BookWinkler03} increasing ordering of $\beta_{1}$ with the energy gaps is unsatisfied for the InP in the considered density range.

Lets turn now to the problem of a Q1D-$hh$ systems, which can be lithographically achieved by placing repulsive electrodes on top of a heterostructure. Following the Chesi \textit{et al.} \cite{Chesi11} approximation, by substituting $\hat k_x\approx 0$,  $\hat k_x^{2}\approx \pi^{2}/L_w^{2}$ and $\hat k_x^{3}\approx 0$ in the Hamiltonian (\ref{eq:HR2D}), for a channel with lateral extent $L_w$ aligned with the $y$ axis, we get
\begin{equation}\label{eq:cheHR1D}
 \hat{\bm H}_{R-1D}^{hh}\approx  -\dfrac{3\pi^{2}}{L_w^{2}}\beta\bm\sigma_y\hat k_y  + \beta\bm\sigma_y\hat k_y^{3},
\end{equation}
which is the expected dependence we want to obtain from our modelling. We simulate the additional constriction by an harmonic potential $\frac{1}{2}m_0\omega^{2}x^{2}$. The same perturbative treatment described above, starting from the Pidgeon-Brown model in the spherical approximation and carried out for the Q2D case, enable us to derive an effective SOI-R  Hamiltonian for Q1D-$hh$ systems

\begin{equation}\label{eq:HR1D}
 \hat{\bm H}_{R-1D}^{hh} = -\alpha\bm\sigma_y\hat k_y + \alpha'\bm\sigma_y\hat k_y^{3}.
\end{equation}

Importantly, a further competitor cubic contribution arises as predicted before \cite{Zhang07,Chesi11}. As a bonus, an appealing concordance with foretold signs is nicely reproduced both for $\hat k_{y}$-linear and $\hat k_{y}$-cubic terms.

The SOI-R  coupling parameters in (\ref{eq:HR1D}) are given by
\begin{equation}\label{eq:alpha-1d}
\alpha_{nn'}=-\dfrac{3i\hbar^{4}\gamma_s^2}{m_0^{2}}
\left(\dfrac{Z^{hl}_{n}\left(I^{hl}_{n}\right)^{2}J^{hl}}{E_{nn'}^{hh}-E_{nn'}^{lh}}- \dfrac{Z^{hs}_{n}\left(I^{hs}_{n}\right)^{2}J^{hs}}{E_{nn'}^{hh}-E_{nn'}^{\revision{ss}}}\right),
\end{equation}
\begin{equation}\label{eq:alphaP-1d}
\alpha'_{nn'}=-\dfrac{3i\hbar^{4}\gamma_s^2}{m_0^{2}}
\left(\dfrac{Z^{hl}_{n}\left(I^{hl}_{n}\right)^{2}K^{hl}}{E_{nn'}^{hh}-E_{nn'}^{lh}}- \dfrac{Z^{hs}_{n}\left(I^{hs}_{n}\right)^{2}K^{hs}}{E_{nn'}^{hh}-E_{nn'}^{\revision{ss}}}\right),
\end{equation}
where the subindex $nn'$ stands for the transversal states, with eigenenergies
\begin{equation}\label{eq:Enn}
 \begin{array}{l}
  E_{nn'}^{hh,lh}=\hbar\omega\sqrt{\gamma_1\pm\gamma_s}\left(n'+\frac{1}{2}\right)+c_{n}\dfrac{(\hbar eF)^{\frac{2}{3}}}{\sqrt[3]{2m^{*}_{hh,lh}}},\\
  E_{nn'}^{\revision{ss}}=\hbar\omega\sqrt{\gamma_1}\left(n'+\frac{1}{2}\right)+c_{n}\dfrac{(\hbar eF)^{\frac{2}{3}}}{\sqrt[3]{2m^{*}_{\revision{ss}}}}+\varDelta_{SO},
 \end{array}
\end{equation}
where $\gamma_1=\gamma_1^{L}-E_p/(3E_g)$.

The coefficients involved in (\ref{eq:alpha-1d})-(\ref{eq:alphaP-1d}) are those in (\ref{eq:InZn}) ---as in the Q2D case--- along with
\begin{eqnarray}\label{eq:Coe}
 J^{hl/hs}=\frac{\sqrt{2}\sqrt{l_{hh}l_{lh,\revision{ss}}^3+l_{hh}^3l_{lh,\revision{ss}}}}{\left(l_{hh}^2+l_{lh,\revision{ss}}^2\right)^{2}},\label{eq:J}\\
 K^{hl/hs}=\sqrt{2}\left(\dfrac{l_{hh}}{l_{lh,\revision{ss}}}+ \dfrac{l_{lh,\revision{ss}}}{l_{hh}}\right)^{-\frac{1}{2}},\label{eq:K}
\end{eqnarray}
where
\begin{equation}\label{eq:eles}
l_{hh,lh}=\sqrt{\frac{\hbar}{m_0\omega}\sqrt{\gamma_1\pm\gamma_s}} \qquad l_{\revision{ss}}=\sqrt{\frac{\hbar}{m_0\omega}\sqrt{\gamma_1}}.
\end{equation}

In additions to the Airy functions (\ref{eq:airy}), transversal eigenstates $\varPsi_{nn'}^{hh,lh,\revision{ss}}(x,z)=\psi_n^{hh,lh,\revision{ss}}(x)\varphi_{n'}^{hh,lh,\revision{ss}}(z)$  comprise the properly normalized Hermite polynomials,
\begin{multline}\label{eq:psi}
 \psi_n^{hh,lh,\revision{ss}}(x)=\frac{1}{\sqrt{2^{n}n!l_{hh,lh,\revision{ss}}\sqrt{\pi}}}\times \\ \times\mathcal{H}_n\left(\frac{x}{l_{hh,lh,\revision{ss}}}\right)\exp\left[-\frac{1}{2}\left(\frac{x}{l_{hh,lh,\revision{ss}}}\right)^{2}\right].
\end{multline}

\begin{figure}
\centering
\includegraphics[width=0.85\linewidth]{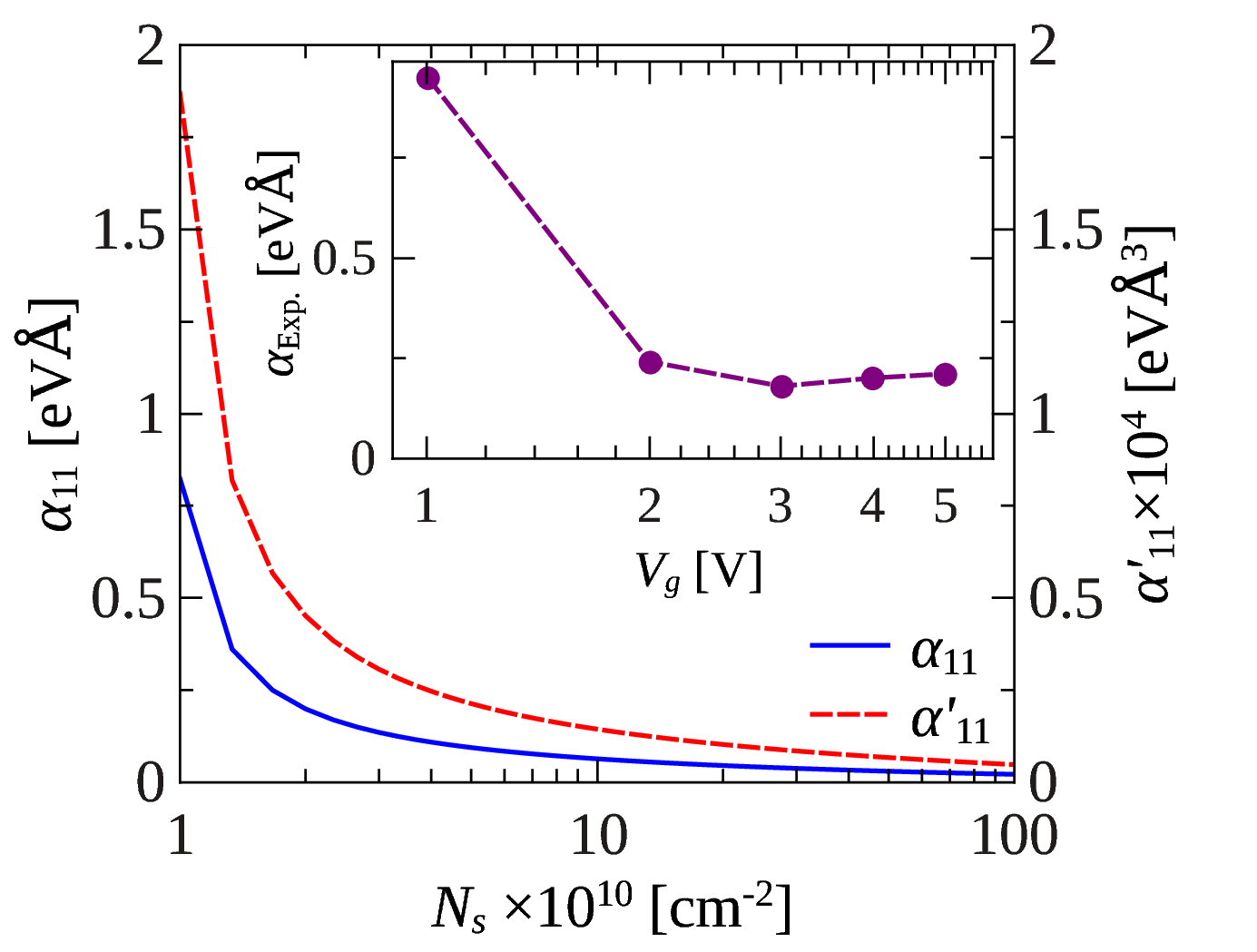}
\caption{(Color online) Q1D coupling parameters $\alpha_{11}$ (\textcolor{blue}{solid line}) and $\alpha'_{11}$ (\textcolor{red}{dashed line}) for GaAs, taken after (\ref{eq:alpha-1d}) and (\ref{eq:alphaP-1d}) respectively. $L_w=l_{hh}=11.8$~nm. \textit{Inset}: SOI-R  strength \mbox{(\textcolor{violet}{-\textbullet-})} measured in 10~nm width holes Ge/Si core/shell nanowires from Ref.~\cite{Hao10}.}
\label{fig:1dexp}
\end{figure}

A similar discussion as the one proposed for the Q2D-$hh$ coupling parameter $\beta$ (\ref{eq:betaQ2D}), is valid here.  Both QD1-$hh$ coupling parameters (\ref{eq:alpha-1d})-(\ref{eq:alphaP-1d}) decrease with an increasing external or intrinsic electric field, as shown in Fig.~\ref{fig:1dexp}. This trend is in qualitative accordance with the SOI-R  strength measured in holes Ge/Si core/shell nanowires \cite{Hao10}, depicted in the inset of Fig.~\ref{fig:1dexp}. We underline that we had even recovered the same order of magnitude.

\begin{figure}
\centering
\includegraphics[width=0.85\linewidth]{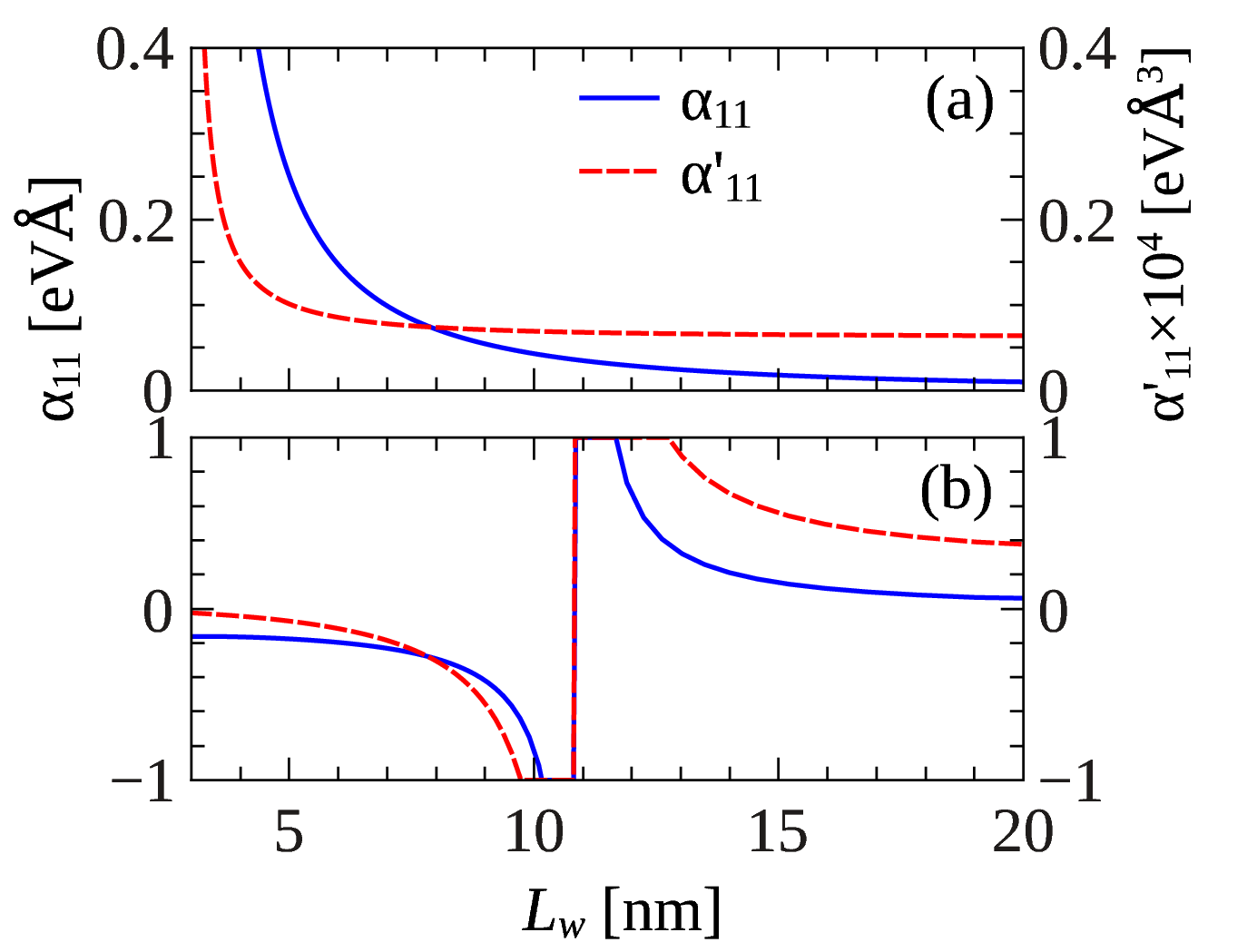}
\caption{(Color online) Q1D coupling parameters $\alpha_{11}$ (\textcolor{blue}{solid line}) and $\alpha'_{11}$ (\textcolor{red}{dashed line}) for GaAs, taken after (\ref{eq:alpha-1d}) and (\ref{eq:alphaP-1d}) respectively, for typical ($N_s>10^{11}$~cm$^{-2}$, panel (a)) and low ($N_s=10^{10}$~cm$^{-2}$, panel (b)) heavy hole densities.}
\label{fig:1DTrends}
\end{figure}

We observe a non-trivial dependence on the characteristic confinement length, $L_w=l_{hh}$ as in (\ref{eq:eles}), which departs from the one in the approximation (\ref{eq:cheHR1D}) \cite{Chesi11}. Indeed, at typical densities ($N_s>10^{11}$~cm$^{-2}$) whose channel extent along $z$ direction is $L_{z} \approx 20$ nm, $\alpha$ and $\alpha'$ decrease whenever the Q1D-$hh$ system become wider, as shown in Fig. \ref{fig:1DTrends} (a). However, $\alpha$ asymptotically tends to zero, while $\alpha'$ promptly turns into a plateau at the vicinity of $\beta$ (\ref{eq:cheHR1D}). This is intuitively clear and straightforward from (\ref{eq:cheHR1D}). In short, our modelling of the Q1D-$hh$ systems recovers the Q2D-$hh$ case, in the limit when $L_w$ grows. On the other hand, at low densities ($N_s=10^{10}$~cm$^{-2}, L_{z} \approx 20$ nm), we found a critical confinement length $L_w\simeq 11$~nm where both coupling parameters exhibit an essential asymptotic discontinuity, as shown in Fig. \ref{fig:1DTrends} (b), which resembles a Fano-like profile.

Owing to a deeper analysis, it seems that this behavior is related to the hole effective mass anisotropy. Indeed, in the spherical approximation, $hh$ effective mass perpendicular to the growth direction ($m^{*}_{hh\bot} = m_0/(\gamma_1+\gamma_s)$) is lighter than $lh$ effective mass  ($m^{*}_{lh\perp} = m_0/(\gamma_1-\gamma_s)$). \revision{Note that the behavior of $\alpha$ and $\alpha'$ is governed by the denominators $E_{nn'}^{hh}-E_{nn'}^{lh}$ and $E_{nn'}^{hh}-E_{nn'}^{\revision{ss}}$ in (\ref{eq:alpha-1d}) and (\ref{eq:alphaP-1d}). The latter hardly vanishes due to the split-off gap $\varDelta_{SO}$. On the other hand, for the first $hh$ and $lh$ subbands ($n=1$ and $n'=0$), following (\ref{eq:Enn}) we have
\begin{align}
 \varDelta\mathcal{E}^{hl} & =E_{1,0}^{hh}-E_{1,0}^{lh},\\
			   & =\dfrac{\hbar\omega}{2}\left(\left(\gamma_1+\gamma_s\right)^{\frac{1}{2}} - \left(\gamma_1-\gamma_s\right)^{\frac{1}{2}} \right) \\
			   & + C_1\left(\dfrac{\left(\hbar eF \right)^{2}}{2m_0} \right)^{\frac{1}{3}}\left(\left(\gamma_1-2\gamma_s\right)^{\frac{1}{3}} - \left(\gamma_1+2\gamma_s\right)^{\frac{1}{3}} \right),\\
			   &=\dfrac{\hbar^{2}}{2m_0L_w^{2}}\varGamma_1-C_1\left(\dfrac{\left(\hbar eF \right)^{2}}{2m_0} \right)^{\frac{1}{3}}\varGamma_2,\label{eq:DEhl}
\end{align}
where $\varGamma_1=\left(\gamma_1+\gamma_s-\left(\gamma_1+\gamma_s\right)^{\frac{1}{2}} \left(\gamma_1-\gamma_s\right)^{\frac{1}{2}} \right)$ and $\varGamma_2=\left(\left(\gamma_1+2\gamma_s\right)^{\frac{1}{3}} - \left(\gamma_1-2\gamma_s\right)^{\frac{1}{3}} \right)$. Both terms in (\ref{eq:DEhl}) are positive definite, thereof there is a value of the characteristic confinement length
\begin{equation}
 L_w^{0}=z^{0}\left( \dfrac{\varGamma_1}{C_1\varGamma_2}\right) ^{\frac{1}{2}},
\end{equation}
that cancels $\varDelta\mathcal{E}^{hl}$. For $L_w=L_w^{0}$, $\alpha$ and $\alpha'$ diverge, if $L_w>L_w^{0}$ are both positive definite while for $L_w<L_w^{0}$ are both negative definite. Despite Shubnikov-de Haas experiments have been performed for hundred-of-nanometers wide quantum wires \cite{Guzenko06}, the beating pattern become considerably weaker for narrow enough systems, where the discontinuity take place ($L_w\approx 10$~nm for $L_z\approx 40$~nm). On the other hand, discontinuities may be detected in magnetic focusing experiments \cite{Chesi11,Nichele14} for QPC patterned in [001]-grown Q2D hole gases, by analyzing the shift of signal peaks as a function of SOI-R strength. As we have shown in Fig.\ref{fig:1DTrends}, the last quantity can be tuned by $L_{w}$. Therefore, for different QPC widths the signal-peak shift mentioned above, may be tailored as well. We foretell an interplay in the broadening of the signal-peak separation, whenever some critical width (alike $L_w^{0}$) is reached, leading then to the opposite tendency for the peaks evolution having trespassed some typical confinement width. The last could support the very existence of such ranges of discontinuity. Nonetheless, in Q1D systems aligned along the growth direction $z$, this discontinuity no longer arises, because of the  effective mass isotropy ($\varDelta\mathcal{E}^{hl}$ never vanishes) in the transversal plane ($xy$), as may be inferred from studies of spectral properties of holes in Q1D systems under SOI-R \cite{Cuan11,Zulicke06,Lu08}.}

In summary, multiband  \textbf{k}$\cdot$\textbf{p} ($8 \times 8$) standard Hamiltonian \cite{Pidgeon66,Efros98}, is a reliable candidate to address the SOI-R coupling modelling for heavy holes embedded in both Q2D and Q1D systems. While we recover the expected form and phenomenology of the effective Hamiltonians \cite{Winkler02,Chesi11}, our accurate SOI-R coupling parameters derivation is confirmed by quantitative and qualitative comparison with experimental measurements. According to our modelling, further estimation procedures of these quantities in specialized semiconductor alloys are no longer needed. As far as Q2D-$hh$ is concerned, we found to be in a remarkable better agreement with respect to previous theoretical simulations. SOI-R coupling parameter anomalous behavior is straightforward understood from our closed analytical expression. We had formally derived an effective SOI-R Hamiltonian whose cubic term is competitor to the linear one. The Q1D-$hh$ coupling parameters also behaves anomalously, and follow recent experimental observations. In this case, we identify two different trends in the dependence with the characteristic confinement length of the nanowire. While at typical densities both coupling parameters decrease as the system become wider, tending to the Q2D case, in the low density regime they exhibit an essential asymptotic discontinuity, that may be produced by permutation of $hh-lh$ effective mass roles.

The authors thank Stefano Chesi and Andreas D. Wieck for useful comments. L.D-C acknowledges the hospitality of the Universidade Federal de S\~{a}o Carlos, Brasil and is grateful to the C\'{a}tedra Itinerante of CLAF-PLAF.

\bibliographystyle{aipnum4-1}
\bibliography{/home/rcuan/Documentos/Field-Work/Literatura/BibDB}

\begin{thebibliography}{35}%
\makeatletter
\providecommand \@ifxundefined [1]{%
 \@ifx{#1\undefined}
}%
\providecommand \@ifnum [1]{%
 \ifnum #1\expandafter \@firstoftwo
 \else \expandafter \@secondoftwo
 \fi
}%
\providecommand \@ifx [1]{%
 \ifx #1\expandafter \@firstoftwo
 \else \expandafter \@secondoftwo
 \fi
}%
\providecommand \natexlab [1]{#1}%
\providecommand \enquote  [1]{``#1''}%
\providecommand \bibnamefont  [1]{#1}%
\providecommand \bibfnamefont [1]{#1}%
\providecommand \citenamefont [1]{#1}%
\providecommand \href@noop [0]{\@secondoftwo}%
\providecommand \href [0]{\begingroup \@sanitize@url \@href}%
\providecommand \@href[1]{\@@startlink{#1}\@@href}%
\providecommand \@@href[1]{\endgroup#1\@@endlink}%
\providecommand \@sanitize@url [0]{\catcode `\\12\catcode `\$12\catcode
  `\&12\catcode `\#12\catcode `\^12\catcode `\_12\catcode `\%12\relax}%
\providecommand \@@startlink[1]{}%
\providecommand \@@endlink[0]{}%
\providecommand \url  [0]{\begingroup\@sanitize@url \@url }%
\providecommand \@url [1]{\endgroup\@href {#1}{\urlprefix }}%
\providecommand \urlprefix  [0]{URL }%
\providecommand \Eprint [0]{\href }%
\providecommand \doibase [0]{http://dx.doi.org/}%
\providecommand \selectlanguage [0]{\@gobble}%
\providecommand \bibinfo  [0]{\@secondoftwo}%
\providecommand \bibfield  [0]{\@secondoftwo}%
\providecommand \translation [1]{[#1]}%
\providecommand \BibitemOpen [0]{}%
\providecommand \bibitemStop [0]{}%
\providecommand \bibitemNoStop [0]{.\EOS\space}%
\providecommand \EOS [0]{\spacefactor3000\relax}%
\providecommand \BibitemShut  [1]{\csname bibitem#1\endcsname}%
\let\auto@bib@innerbib\@empty
%</preamble>
\bibitem [{\citenamefont {Awschalom}\ and\ \citenamefont
  {Flatt{\'e}}(2007)}]{Awschalom07}%
  \BibitemOpen
  \bibfield  {author} {\bibinfo {author} {\bibfnamefont {D.~D.}\ \bibnamefont
  {Awschalom}}\ and\ \bibinfo {author} {\bibfnamefont {M.~E.}\ \bibnamefont
  {Flatt{\'e}}},\ }\href@noop {} {\bibfield  {journal} {\bibinfo  {journal}
  {Nature Phys.}\ }\textbf {\bibinfo {volume} {3}},\ \bibinfo {pages} {153}
  (\bibinfo {year} {2007})}\BibitemShut {NoStop}%
\bibitem [{\citenamefont {Bychkov}\ and\ \citenamefont
  {Rashba}(1984{\natexlab{a}})}]{Rashba84a}%
  \BibitemOpen
  \bibfield  {author} {\bibinfo {author} {\bibfnamefont {Y.~A.}\ \bibnamefont
  {Bychkov}}\ and\ \bibinfo {author} {\bibfnamefont {E.~I.}\ \bibnamefont
  {Rashba}},\ }\href@noop {} {\bibfield  {journal} {\bibinfo  {journal} {J.
  Phys. C: Solid State Phys.}\ }\textbf {\bibinfo {volume} {17}},\ \bibinfo
  {pages} {6039} (\bibinfo {year} {1984}{\natexlab{a}})}\BibitemShut {NoStop}%
\bibitem [{\citenamefont {Bychkov}\ and\ \citenamefont
  {Rashba}(1984{\natexlab{b}})}]{Rashba84b}%
  \BibitemOpen
  \bibfield  {author} {\bibinfo {author} {\bibfnamefont {Y.~A.}\ \bibnamefont
  {Bychkov}}\ and\ \bibinfo {author} {\bibfnamefont {E.~I.}\ \bibnamefont
  {Rashba}},\ }\href@noop {} {\bibfield  {journal} {\bibinfo  {journal} {JETP
  Lett.}\ }\textbf {\bibinfo {volume} {39}},\ \bibinfo {pages} {2} (\bibinfo
  {year} {1984}{\natexlab{b}})}\BibitemShut {NoStop}%
\bibitem [{\citenamefont {Datta}\ and\ \citenamefont {Das}(1990)}]{Datta90}%
  \BibitemOpen
  \bibfield  {author} {\bibinfo {author} {\bibfnamefont {S.}~\bibnamefont
  {Datta}}\ and\ \bibinfo {author} {\bibfnamefont {B.}~\bibnamefont {Das}},\
  }\href@noop {} {\bibfield  {journal} {\bibinfo  {journal} {Appl. Phys.
  Lett.}\ }\textbf {\bibinfo {volume} {56}},\ \bibinfo {pages} {665} (\bibinfo
  {year} {1990})}\BibitemShut {NoStop}%
\bibitem [{\citenamefont {Pala}\ \emph {et~al.}(2004)\citenamefont {Pala},
  \citenamefont {Governale}, \citenamefont {K{\"o}nig}, \citenamefont
  {Z{\"u}licke},\ and\ \citenamefont {Iannaccone}}]{Pala04}%
  \BibitemOpen
  \bibfield  {author} {\bibinfo {author} {\bibfnamefont {M.~G.}\ \bibnamefont
  {Pala}}, \bibinfo {author} {\bibfnamefont {M.}~\bibnamefont {Governale}},
  \bibinfo {author} {\bibfnamefont {J.}~\bibnamefont {K{\"o}nig}}, \bibinfo
  {author} {\bibfnamefont {U.}~\bibnamefont {Z{\"u}licke}}, \ and\ \bibinfo
  {author} {\bibfnamefont {G.}~\bibnamefont {Iannaccone}},\ }\href@noop {}
  {\bibfield  {journal} {\bibinfo  {journal} {Phys. Rev. B}\ }\textbf {\bibinfo
  {volume} {69}},\ \bibinfo {pages} {045304} (\bibinfo {year}
  {2004})}\BibitemShut {NoStop}%
\bibitem [{\citenamefont {Shelykh}, \citenamefont {Galkin},\ and\ \citenamefont
  {Bagraev}(2005)}]{Shelykh05}%
  \BibitemOpen
  \bibfield  {author} {\bibinfo {author} {\bibfnamefont {I.~A.}\ \bibnamefont
  {Shelykh}}, \bibinfo {author} {\bibfnamefont {N.~G.}\ \bibnamefont {Galkin}},
  \ and\ \bibinfo {author} {\bibfnamefont {N.~T.}\ \bibnamefont {Bagraev}},\
  }\href@noop {} {\bibfield  {journal} {\bibinfo  {journal} {Phys. Rev. B}\
  }\textbf {\bibinfo {volume} {72}},\ \bibinfo {pages} {235316} (\bibinfo
  {year} {2005})}\BibitemShut {NoStop}%
\bibitem [{\citenamefont {Diago-Cisneros}\ and\ \citenamefont
  {Mireles}(2013)}]{Leo13}%
  \BibitemOpen
  \bibfield  {author} {\bibinfo {author} {\bibfnamefont {L.}~\bibnamefont
  {Diago-Cisneros}}\ and\ \bibinfo {author} {\bibfnamefont {F.}~\bibnamefont
  {Mireles}},\ }\href@noop {} {\bibfield  {journal} {\bibinfo  {journal} {J.
  Appl. Phys}\ }\textbf {\bibinfo {volume} {114}},\ \bibinfo {pages} {193706}
  (\bibinfo {year} {2013})}\BibitemShut {NoStop}%
\bibitem [{\citenamefont {Wieck}\ \emph {et~al.}(1984)\citenamefont {Wieck},
  \citenamefont {Batke}, \citenamefont {Heitmann}, \citenamefont {Kotthaus},\
  and\ \citenamefont {Bangert}}]{Wieck84}%
  \BibitemOpen
  \bibfield  {author} {\bibinfo {author} {\bibfnamefont {A.~D.}\ \bibnamefont
  {Wieck}}, \bibinfo {author} {\bibfnamefont {E.}~\bibnamefont {Batke}},
  \bibinfo {author} {\bibfnamefont {D.}~\bibnamefont {Heitmann}}, \bibinfo
  {author} {\bibfnamefont {J.~P.}\ \bibnamefont {Kotthaus}}, \ and\ \bibinfo
  {author} {\bibfnamefont {E.}~\bibnamefont {Bangert}},\ }\href@noop {}
  {\bibfield  {journal} {\bibinfo  {journal} {Phys. Rev. Lett.}\ }\textbf
  {\bibinfo {volume} {53}},\ \bibinfo {pages} {493} (\bibinfo {year}
  {1984})}\BibitemShut {NoStop}%
\bibitem [{\citenamefont {Gerchikov}\ and\ \citenamefont
  {Subashiev}(1992)}]{Gerchikov92}%
  \BibitemOpen
  \bibfield  {author} {\bibinfo {author} {\bibfnamefont {L.~G.}\ \bibnamefont
  {Gerchikov}}\ and\ \bibinfo {author} {\bibfnamefont {A.~V.}\ \bibnamefont
  {Subashiev}},\ }\href@noop {} {\bibfield  {journal} {\bibinfo  {journal}
  {Sov. Phys. Semicond.}\ }\textbf {\bibinfo {volume} {26}},\ \bibinfo {pages}
  {73} (\bibinfo {year} {1992})}\BibitemShut {NoStop}%
\bibitem [{\citenamefont {Winkler}\ \emph {et~al.}(2002)\citenamefont
  {Winkler}, \citenamefont {Noh}, \citenamefont {Tutuc},\ and\ \citenamefont
  {Shayegan}}]{Winkler02}%
  \BibitemOpen
  \bibfield  {author} {\bibinfo {author} {\bibfnamefont {R.}~\bibnamefont
  {Winkler}}, \bibinfo {author} {\bibfnamefont {H.}~\bibnamefont {Noh}},
  \bibinfo {author} {\bibfnamefont {E.}~\bibnamefont {Tutuc}}, \ and\ \bibinfo
  {author} {\bibfnamefont {M.}~\bibnamefont {Shayegan}},\ }\href@noop {}
  {\bibfield  {journal} {\bibinfo  {journal} {Phys. Rev. B}\ }\textbf {\bibinfo
  {volume} {65}},\ \bibinfo {pages} {155303} (\bibinfo {year}
  {2002})}\BibitemShut {NoStop}%
\bibitem [{\citenamefont {Winkler}()}]{BookWinkler03}%
  \BibitemOpen
  \bibfield  {author} {\bibinfo {author} {\bibfnamefont {R.}~\bibnamefont
  {Winkler}},\ }\href@noop {} {\emph {\bibinfo {title} {{Spin-orbit coupling
  effects in two-dimensional electron and hole systems}}}}\ (\bibinfo
  {publisher} {Springer-Verlag},\ \bibinfo {address} {Berlin Heidelberg,
  2003})\BibitemShut {NoStop}%
\bibitem [{\citenamefont {Habib}, \citenamefont {Shayegan},\ and\ \citenamefont
  {Winkler}(2009)}]{Habib09}%
  \BibitemOpen
  \bibfield  {author} {\bibinfo {author} {\bibfnamefont {B.}~\bibnamefont
  {Habib}}, \bibinfo {author} {\bibfnamefont {M.}~\bibnamefont {Shayegan}}, \
  and\ \bibinfo {author} {\bibfnamefont {R.}~\bibnamefont {Winkler}},\
  }\href@noop {} {\bibfield  {journal} {\bibinfo  {journal} {Semicond. Sci.
  Technol.}\ }\textbf {\bibinfo {volume} {23}},\ \bibinfo {pages} {064002}
  (\bibinfo {year} {2009})}\BibitemShut {NoStop}%
\bibitem [{\citenamefont {{de Andrada e Silva}}, \citenamefont {{La Rocca}},\
  and\ \citenamefont {Bassani}(1994)}]{Andrada94}%
  \BibitemOpen
  \bibfield  {author} {\bibinfo {author} {\bibfnamefont {E.~A.}\ \bibnamefont
  {{de Andrada e Silva}}}, \bibinfo {author} {\bibfnamefont {G.~C.}\
  \bibnamefont {{La Rocca}}}, \ and\ \bibinfo {author} {\bibfnamefont
  {F.}~\bibnamefont {Bassani}},\ }\href@noop {} {\bibfield  {journal} {\bibinfo
   {journal} {Phys. Rev. B}\ }\textbf {\bibinfo {volume} {50}},\ \bibinfo
  {pages} {8523} (\bibinfo {year} {1994})}\BibitemShut {NoStop}%
\bibitem [{\citenamefont {Quay}\ \emph {et~al.}(2010)\citenamefont {Quay},
  \citenamefont {Hughes}, \citenamefont {Sulpizio}, \citenamefont {Pfeffer},
  \citenamefont {Baldwin}, \citenamefont {West}, \citenamefont
  {Goldhaber-Gordon},\ and\ \citenamefont {de~Picciotto}}]{Quay10}%
  \BibitemOpen
  \bibfield  {author} {\bibinfo {author} {\bibfnamefont {C.~H.~L.}\
  \bibnamefont {Quay}}, \bibinfo {author} {\bibfnamefont {T.~L.}\ \bibnamefont
  {Hughes}}, \bibinfo {author} {\bibfnamefont {J.~A.}\ \bibnamefont
  {Sulpizio}}, \bibinfo {author} {\bibfnamefont {L.~N.}\ \bibnamefont
  {Pfeffer}}, \bibinfo {author} {\bibfnamefont {K.~W.}\ \bibnamefont
  {Baldwin}}, \bibinfo {author} {\bibfnamefont {K.~W.}\ \bibnamefont {West}},
  \bibinfo {author} {\bibfnamefont {D.}~\bibnamefont {Goldhaber-Gordon}}, \
  and\ \bibinfo {author} {\bibfnamefont {R.}~\bibnamefont {de~Picciotto}},\
  }\href@noop {} {\bibfield  {journal} {\bibinfo  {journal} {Nature Phys.}\
  }\textbf {\bibinfo {volume} {6}},\ \bibinfo {pages} {336} (\bibinfo {year}
  {2010})}\BibitemShut {NoStop}%
\bibitem [{\citenamefont {Governale}\ and\ \citenamefont
  {Z{\"u}licke}(2003)}]{Governale03}%
  \BibitemOpen
  \bibfield  {author} {\bibinfo {author} {\bibfnamefont {M.}~\bibnamefont
  {Governale}}\ and\ \bibinfo {author} {\bibfnamefont {U.}~\bibnamefont
  {Z{\"u}licke}},\ }\href@noop {} {\bibfield  {journal} {\bibinfo  {journal}
  {J. Supercond. Nov. Magn.}\ }\textbf {\bibinfo {volume} {16}},\ \bibinfo
  {pages} {257} (\bibinfo {year} {2003})}\BibitemShut {NoStop}%
\bibitem [{\citenamefont {Zhang}\ and\ \citenamefont {Xia}(2007)}]{Zhang07}%
  \BibitemOpen
  \bibfield  {author} {\bibinfo {author} {\bibfnamefont {X.~W.}\ \bibnamefont
  {Zhang}}\ and\ \bibinfo {author} {\bibfnamefont {J.~B.}\ \bibnamefont
  {Xia}},\ }\href@noop {} {\bibfield  {journal} {\bibinfo  {journal} {J. Phys.
  D: Appl. Phys.}\ }\textbf {\bibinfo {volume} {40}},\ \bibinfo {pages} {541}
  (\bibinfo {year} {2007})}\BibitemShut {NoStop}%
\bibitem [{\citenamefont {Cuan}\ and\ \citenamefont
  {Diago-Cisneros}(2011)}]{Cuan11}%
  \BibitemOpen
  \bibfield  {author} {\bibinfo {author} {\bibfnamefont {R.}~\bibnamefont
  {Cuan}}\ and\ \bibinfo {author} {\bibfnamefont {L.}~\bibnamefont
  {Diago-Cisneros}},\ }\href@noop {} {\bibfield  {journal} {\bibinfo  {journal}
  {J. Appl. Phys}\ }\textbf {\bibinfo {volume} {110}},\ \bibinfo {pages}
  {113705} (\bibinfo {year} {2011})}\BibitemShut {NoStop}%
\bibitem [{\citenamefont {Chesi}\ \emph {et~al.}(2011)\citenamefont {Chesi},
  \citenamefont {Giuliani}, \citenamefont {Rokhinson}, \citenamefont
  {Pfeffer},\ and\ \citenamefont {West}}]{Chesi11}%
  \BibitemOpen
  \bibfield  {author} {\bibinfo {author} {\bibfnamefont {S.}~\bibnamefont
  {Chesi}}, \bibinfo {author} {\bibfnamefont {G.~F.}\ \bibnamefont {Giuliani}},
  \bibinfo {author} {\bibfnamefont {L.~P.}\ \bibnamefont {Rokhinson}}, \bibinfo
  {author} {\bibfnamefont {L.~N.}\ \bibnamefont {Pfeffer}}, \ and\ \bibinfo
  {author} {\bibfnamefont {K.~W.}\ \bibnamefont {West}},\ }\href@noop {}
  {\bibfield  {journal} {\bibinfo  {journal} {Phys. Rev. Lett.}\ }\textbf
  {\bibinfo {volume} {106}},\ \bibinfo {pages} {236601} (\bibinfo {year}
  {2011})}\BibitemShut {NoStop}%
\bibitem [{\citenamefont {Hao}\ \emph {et~al.}(2010)\citenamefont {Hao},
  \citenamefont {Tu}, \citenamefont {Cao}, \citenamefont {Zhou}, \citenamefont
  {Li}, \citenamefont {Guo}, \citenamefont {Fung}, \citenamefont {Ji},
  \citenamefont {Guo},\ and\ \citenamefont {Lu}}]{Hao10}%
  \BibitemOpen
  \bibfield  {author} {\bibinfo {author} {\bibfnamefont {X.~J.}\ \bibnamefont
  {Hao}}, \bibinfo {author} {\bibfnamefont {T.}~\bibnamefont {Tu}}, \bibinfo
  {author} {\bibfnamefont {G.}~\bibnamefont {Cao}}, \bibinfo {author}
  {\bibfnamefont {C.}~\bibnamefont {Zhou}}, \bibinfo {author} {\bibfnamefont
  {H.~O.}\ \bibnamefont {Li}}, \bibinfo {author} {\bibfnamefont {G.~C.}\
  \bibnamefont {Guo}}, \bibinfo {author} {\bibfnamefont {W.~Y.}\ \bibnamefont
  {Fung}}, \bibinfo {author} {\bibfnamefont {Z.}~\bibnamefont {Ji}}, \bibinfo
  {author} {\bibfnamefont {G.~P.}\ \bibnamefont {Guo}}, \ and\ \bibinfo
  {author} {\bibfnamefont {W.}~\bibnamefont {Lu}},\ }\href@noop {} {\bibfield
  {journal} {\bibinfo  {journal} {Nano Letters}\ }\textbf {\bibinfo {volume}
  {10}},\ \bibinfo {pages} {2956} (\bibinfo {year} {2010})}\BibitemShut
  {NoStop}%
\bibitem [{\citenamefont {Rashba}\ and\ \citenamefont
  {Sherman}(1988)}]{Rashba88}%
  \BibitemOpen
  \bibfield  {author} {\bibinfo {author} {\bibfnamefont {E.~I.}\ \bibnamefont
  {Rashba}}\ and\ \bibinfo {author} {\bibfnamefont {E.~Y.}\ \bibnamefont
  {Sherman}},\ }\href@noop {} {\bibfield  {journal} {\bibinfo  {journal} {Phys.
  Lett. A}\ }\textbf {\bibinfo {volume} {129}},\ \bibinfo {pages} {175}
  (\bibinfo {year} {1988})}\BibitemShut {NoStop}%
\bibitem [{\citenamefont {Nichele}\ \emph {et~al.}(2014)\citenamefont
  {Nichele}, \citenamefont {Chesi}, \citenamefont {Hennel}, \citenamefont
  {Wittmann}, \citenamefont {Gerl}, \citenamefont {Wegscheider}, \citenamefont
  {Loss}, \citenamefont {Ihn},\ and\ \citenamefont {Ensslin}}]{Nichele14}%
  \BibitemOpen
  \bibfield  {author} {\bibinfo {author} {\bibfnamefont {F.}~\bibnamefont
  {Nichele}}, \bibinfo {author} {\bibfnamefont {S.}~\bibnamefont {Chesi}},
  \bibinfo {author} {\bibfnamefont {S.}~\bibnamefont {Hennel}}, \bibinfo
  {author} {\bibfnamefont {A.}~\bibnamefont {Wittmann}}, \bibinfo {author}
  {\bibfnamefont {C.}~\bibnamefont {Gerl}}, \bibinfo {author} {\bibfnamefont
  {W.}~\bibnamefont {Wegscheider}}, \bibinfo {author} {\bibfnamefont
  {D.}~\bibnamefont {Loss}}, \bibinfo {author} {\bibfnamefont {T.}~\bibnamefont
  {Ihn}}, \ and\ \bibinfo {author} {\bibfnamefont {K.}~\bibnamefont
  {Ensslin}},\ }\href@noop {} {\bibfield  {journal} {\bibinfo  {journal} {Phys.
  Rev. Lett.}\ }\textbf {\bibinfo {volume} {113}},\ \bibinfo {pages} {046801}
  (\bibinfo {year} {2014})}\BibitemShut {NoStop}%
\bibitem [{\citenamefont {de~Andrada~e Silva}\ and\ \citenamefont
  {la~Rocca}(2003)}]{Andrada03}%
  \BibitemOpen
  \bibfield  {author} {\bibinfo {author} {\bibfnamefont {E.~A.}\ \bibnamefont
  {de~Andrada~e Silva}}\ and\ \bibinfo {author} {\bibfnamefont {G.~C.}\
  \bibnamefont {la~Rocca}},\ }\href@noop {} {\bibfield  {journal} {\bibinfo
  {journal} {Phys. Rev. B}\ }\textbf {\bibinfo {volume} {67}},\ \bibinfo
  {pages} {165318} (\bibinfo {year} {2003})}\BibitemShut {NoStop}%
\bibitem [{\citenamefont {Guzenko}\ \emph {et~al.}(2006)\citenamefont
  {Guzenko}, \citenamefont {Knobbe}, \citenamefont {Hardtdegen}, \citenamefont
  {Sch{\"a}pers},\ and\ \citenamefont {Bringer}}]{Guzenko06}%
  \BibitemOpen
  \bibfield  {author} {\bibinfo {author} {\bibfnamefont {V.~A.}\ \bibnamefont
  {Guzenko}}, \bibinfo {author} {\bibfnamefont {J.}~\bibnamefont {Knobbe}},
  \bibinfo {author} {\bibfnamefont {H.}~\bibnamefont {Hardtdegen}}, \bibinfo
  {author} {\bibfnamefont {T.}~\bibnamefont {Sch{\"a}pers}}, \ and\ \bibinfo
  {author} {\bibfnamefont {A.}~\bibnamefont {Bringer}},\ }\href@noop {}
  {\bibfield  {journal} {\bibinfo  {journal} {Appl. Phys. Lett.}\ }\textbf
  {\bibinfo {volume} {88}},\ \bibinfo {pages} {032102} (\bibinfo {year}
  {2006})}\BibitemShut {NoStop}%
\bibitem [{\citenamefont {Pidgeon}\ and\ \citenamefont
  {Brown}(1966)}]{Pidgeon66}%
  \BibitemOpen
  \bibfield  {author} {\bibinfo {author} {\bibfnamefont {C.~R.}\ \bibnamefont
  {Pidgeon}}\ and\ \bibinfo {author} {\bibfnamefont {N.~R.}\ \bibnamefont
  {Brown}},\ }\href@noop {} {\bibfield  {journal} {\bibinfo  {journal} {Phys.
  Rev.}\ }\textbf {\bibinfo {volume} {146}},\ \bibinfo {pages} {575} (\bibinfo
  {year} {1966})}\BibitemShut {NoStop}%
\bibitem [{\citenamefont {Efros}\ and\ \citenamefont {Rosen}(1998)}]{Efros98}%
  \BibitemOpen
  \bibfield  {author} {\bibinfo {author} {\bibfnamefont {A.~L.}\ \bibnamefont
  {Efros}}\ and\ \bibinfo {author} {\bibfnamefont {M.}~\bibnamefont {Rosen}},\
  }\href@noop {} {\bibfield  {journal} {\bibinfo  {journal} {Phys. Rev. B}\
  }\textbf {\bibinfo {volume} {58}},\ \bibinfo {pages} {7120} (\bibinfo {year}
  {1998})}\BibitemShut {NoStop}%
\bibitem [{Note1()}]{Note1}%
  \BibitemOpen
  \bibinfo {note} {{Here $\protect \mathaccentV {hat}05E{\protect \bm
  {H}}_{\protect \text {PB}}$ is the Pidgeon-Brown Hamiltonian while $(\protect
  \mathbb {I/O})_{8}$ stands for the eight-dimension
  identity-matrix/null-vector, respectively.}}\BibitemShut {Stop}%
\bibitem [{\citenamefont {L{\"o}wdin}(1951)}]{Lowdin51}%
  \BibitemOpen
  \bibfield  {author} {\bibinfo {author} {\bibfnamefont {P.~O.}\ \bibnamefont
  {L{\"o}wdin}},\ }\href@noop {} {\bibfield  {journal} {\bibinfo  {journal} {J.
  Chem. Phys.}\ }\textbf {\bibinfo {volume} {19}},\ \bibinfo {pages} {1396}
  (\bibinfo {year} {1951})}\BibitemShut {NoStop}%
\bibitem [{\citenamefont {Grbi{\'c}}\ \emph {et~al.}(2005)\citenamefont
  {Grbi{\'c}}, \citenamefont {Ellenberger}, \citenamefont {Ihn}, \citenamefont
  {Ensslin}, \citenamefont {Reuter},\ and\ \citenamefont {Wieck}}]{Grbic05}%
  \BibitemOpen
  \bibfield  {author} {\bibinfo {author} {\bibfnamefont {B.}~\bibnamefont
  {Grbi{\'c}}}, \bibinfo {author} {\bibfnamefont {C.}~\bibnamefont
  {Ellenberger}}, \bibinfo {author} {\bibfnamefont {T.}~\bibnamefont {Ihn}},
  \bibinfo {author} {\bibfnamefont {K.}~\bibnamefont {Ensslin}}, \bibinfo
  {author} {\bibfnamefont {D.}~\bibnamefont {Reuter}}, \ and\ \bibinfo {author}
  {\bibfnamefont {A.~D.}\ \bibnamefont {Wieck}},\ }\href@noop {} {\bibfield
  {journal} {\bibinfo  {journal} {AIP Conf. Proc.}\ }\textbf {\bibinfo {volume}
  {772}},\ \bibinfo {pages} {407} (\bibinfo {year} {2005})}\BibitemShut
  {NoStop}%
\bibitem [{\citenamefont {Grbi{\'c}}(2007)}]{GrbicPhD07}%
  \BibitemOpen
  \bibfield  {author} {\bibinfo {author} {\bibfnamefont {B.}~\bibnamefont
  {Grbi{\'c}}},\ }\emph {\bibinfo {title} {{Hole transport and spin-orbit
  coupling in p-type GaAs nanostructures}}},\ \href@noop {} {Ph.D. thesis},\
  \bibinfo  {school} {ETH Zurich} (\bibinfo {year} {2007})\BibitemShut
  {NoStop}%
\bibitem [{\citenamefont {Grbi{\'c}}\ \emph
  {et~al.}(2008{\natexlab{a}})\citenamefont {Grbi{\'c}}, \citenamefont
  {Leturcq}, \citenamefont {Ihn}, \citenamefont {Ensslin}, \citenamefont
  {Reuter},\ and\ \citenamefont {Wieck}}]{Grbic08}%
  \BibitemOpen
  \bibfield  {author} {\bibinfo {author} {\bibfnamefont {B.}~\bibnamefont
  {Grbi{\'c}}}, \bibinfo {author} {\bibfnamefont {R.}~\bibnamefont {Leturcq}},
  \bibinfo {author} {\bibfnamefont {T.}~\bibnamefont {Ihn}}, \bibinfo {author}
  {\bibfnamefont {K.}~\bibnamefont {Ensslin}}, \bibinfo {author} {\bibfnamefont
  {D.}~\bibnamefont {Reuter}}, \ and\ \bibinfo {author} {\bibfnamefont {A.~D.}\
  \bibnamefont {Wieck}},\ }\href@noop {} {\bibfield  {journal} {\bibinfo
  {journal} {Physica E}\ }\textbf {\bibinfo {volume} {40}},\ \bibinfo {pages}
  {2144} (\bibinfo {year} {2008}{\natexlab{a}})}\BibitemShut {NoStop}%
\bibitem [{\citenamefont {Grbi{\'c}}\ \emph
  {et~al.}(2008{\natexlab{b}})\citenamefont {Grbi{\'c}}, \citenamefont
  {Leturcq}, \citenamefont {Ihn}, \citenamefont {Ensslin}, \citenamefont
  {Reuter},\ and\ \citenamefont {Wieck}}]{Grbic08a}%
  \BibitemOpen
  \bibfield  {author} {\bibinfo {author} {\bibfnamefont {B.}~\bibnamefont
  {Grbi{\'c}}}, \bibinfo {author} {\bibfnamefont {R.}~\bibnamefont {Leturcq}},
  \bibinfo {author} {\bibfnamefont {T.}~\bibnamefont {Ihn}}, \bibinfo {author}
  {\bibfnamefont {K.}~\bibnamefont {Ensslin}}, \bibinfo {author} {\bibfnamefont
  {D.}~\bibnamefont {Reuter}}, \ and\ \bibinfo {author} {\bibfnamefont {A.~D.}\
  \bibnamefont {Wieck}},\ }\href@noop {} {\bibfield  {journal} {\bibinfo
  {journal} {Phys. Rev. B}\ }\textbf {\bibinfo {volume} {77}},\ \bibinfo
  {pages} {125312} (\bibinfo {year} {2008}{\natexlab{b}})}\BibitemShut
  {NoStop}%
\bibitem [{\citenamefont {Lu}\ \emph {et~al.}(1998)\citenamefont {Lu},
  \citenamefont {Yau}, \citenamefont {Shukla}, \citenamefont {Shayegan},
  \citenamefont {Wissinger}, \citenamefont {R{\"o}ssler},\ and\ \citenamefont
  {Winkler}}]{Lu98}%
  \BibitemOpen
  \bibfield  {author} {\bibinfo {author} {\bibfnamefont {J.~P.}\ \bibnamefont
  {Lu}}, \bibinfo {author} {\bibfnamefont {J.~B.}\ \bibnamefont {Yau}},
  \bibinfo {author} {\bibfnamefont {S.~P.}\ \bibnamefont {Shukla}}, \bibinfo
  {author} {\bibfnamefont {M.}~\bibnamefont {Shayegan}}, \bibinfo {author}
  {\bibfnamefont {L.}~\bibnamefont {Wissinger}}, \bibinfo {author}
  {\bibfnamefont {U.}~\bibnamefont {R{\"o}ssler}}, \ and\ \bibinfo {author}
  {\bibfnamefont {R.}~\bibnamefont {Winkler}},\ }\href@noop {} {\bibfield
  {journal} {\bibinfo  {journal} {Phys. Rev. Lett.}\ }\textbf {\bibinfo
  {volume} {81}},\ \bibinfo {pages} {1282} (\bibinfo {year}
  {1998})}\BibitemShut {NoStop}%
\bibitem [{\citenamefont {Grbi{\'c}}\ \emph {et~al.}(2004)\citenamefont
  {Grbi{\'c}}, \citenamefont {Ellenberger}, \citenamefont {Ihn}, \citenamefont
  {Ensslin}, \citenamefont {Reuter},\ and\ \citenamefont {Wieck}}]{Grbic04}%
  \BibitemOpen
  \bibfield  {author} {\bibinfo {author} {\bibfnamefont {B.}~\bibnamefont
  {Grbi{\'c}}}, \bibinfo {author} {\bibfnamefont {C.}~\bibnamefont
  {Ellenberger}}, \bibinfo {author} {\bibfnamefont {T.}~\bibnamefont {Ihn}},
  \bibinfo {author} {\bibfnamefont {K.}~\bibnamefont {Ensslin}}, \bibinfo
  {author} {\bibfnamefont {D.}~\bibnamefont {Reuter}}, \ and\ \bibinfo {author}
  {\bibfnamefont {A.~D.}\ \bibnamefont {Wieck}},\ }\href@noop {} {\bibfield
  {journal} {\bibinfo  {journal} {Appl. Phys. Lett.}\ }\textbf {\bibinfo
  {volume} {85}},\ \bibinfo {pages} {2277} (\bibinfo {year}
  {2004})}\BibitemShut {NoStop}%
\bibitem [{\citenamefont {Z{\"u}licke}(2006)}]{Zulicke06}%
  \BibitemOpen
  \bibfield  {author} {\bibinfo {author} {\bibfnamefont {U.}~\bibnamefont
  {Z{\"u}licke}},\ }\href@noop {} {\bibfield  {journal} {\bibinfo  {journal}
  {Phys. Stat. Sol. (c)}\ }\textbf {\bibinfo {volume} {3}},\ \bibinfo {pages}
  {4354} (\bibinfo {year} {2006})}\BibitemShut {NoStop}%
\bibitem [{\citenamefont {L{\"u}}, \citenamefont {Z{\"u}licke},\ and\
  \citenamefont {Wu}(2008)}]{Lu08}%
  \BibitemOpen
  \bibfield  {author} {\bibinfo {author} {\bibfnamefont {C.}~\bibnamefont
  {L{\"u}}}, \bibinfo {author} {\bibfnamefont {U.}~\bibnamefont {Z{\"u}licke}},
  \ and\ \bibinfo {author} {\bibfnamefont {M.~W.}\ \bibnamefont {Wu}},\
  }\href@noop {} {\bibfield  {journal} {\bibinfo  {journal} {Phys. Rev. B}\
  }\textbf {\bibinfo {volume} {78}},\ \bibinfo {pages} {165321} (\bibinfo
  {year} {2008})}\BibitemShut {NoStop}%
\end{thebibliography}%

\end{document}